\documentclass[aps,pra,reprint,amsmath,amssymb,superscriptaddress,nobibnotes]{revtex4-2}
\usepackage[utf8]{inputenc}
\usepackage[english]{babel}
\usepackage[T1]{fontenc}
\usepackage{natbib}
\usepackage[usenames,dvipsnames]{color}
\usepackage{color}
\usepackage{xcolor}
\usepackage{mathtools}
\usepackage{graphicx}
\usepackage{amssymb}
\usepackage{amsmath}
\usepackage{amsthm}
\usepackage{amsfonts}
\usepackage{braket}
\usepackage{soul}
\usepackage{subfigure}
\usepackage{booktabs}
\usepackage{hyperref}
\newtheorem{theorem}{Theorem}

\usepackage{braket}
\usepackage{amsfonts}
\usepackage{dsfont}
\usepackage{color}
\usepackage[lmargin = 25mm, rmargin = 25mm, tmargin = 15mm]{geometry}
\usepackage{subfigure}
\usepackage{graphicx}
\usepackage{bibentry}

\definecolor{CCTLABgreen}{RGB}{0,128,0}

\begin{document}



\title{Nondestructive discrimination of Bell states between distant parties}

\author{Bohdan Bilash}
\affiliation{Center for Quantum Information, Korea Institute of Science and Technology (KIST), Seoul, 02792, Republic of Korea}
\affiliation{Division of Nano \& Information Technology, KIST School, Korea University of Science and Technology, Seoul 02792, Republic of Korea}

\author{Youngrong Lim}
\affiliation{School of Computational Sciences, Korea Institute for Advanced Study, Seoul 02455, Republic of Korea}

\author{Hyukjoon Kwon}
\affiliation{School of Computational Sciences, Korea Institute for Advanced Study, Seoul 02455, Republic of Korea}

\author{Yosep Kim}
\affiliation{Center for Quantum Information, Korea Institute of Science and Technology (KIST), Seoul, 02792, Republic of Korea}
\affiliation{Department of Physics, Korea University, Seoul, 02841, Republic of Korea}

\author{Hyang-Tag Lim}
\affiliation{Center for Quantum Information, Korea Institute of Science and Technology (KIST), Seoul, 02792, Republic of Korea}
\affiliation{Division of Nano \& Information Technology, KIST School, Korea University of Science and Technology, Seoul 02792, Republic of Korea}


\author{Wooyeong Song}
\email{wysong@kisti.re.kr}
\affiliation{Center for Quantum Information, Korea Institute of Science and Technology (KIST), Seoul, 02792, Republic of Korea}
\affiliation{Quantum Network Research Center, Korea Institute of Science and Technology Information (KISTI), Daejeon, 34141, Republic of Korea}

\author{Yong-Su Kim}
\email{yong-su.kim@kist.re.kr}
\affiliation{Center for Quantum Information, Korea Institute of Science and Technology (KIST), Seoul, 02792, Republic of Korea}
\affiliation{Division of Nano \& Information Technology, KIST School, Korea University of Science and Technology, Seoul 02792, Republic of Korea}

\begin{abstract}
Identifying Bell states without destroying it is frequently dealt with in nowadays quantum technologies such as quantum communication and quantum computing. In practice, quantum entangled states are often distributed among distant parties, and it might be required to determine them separately at each location, without inline communication between parties. We present a scheme for discriminating an arbitrary Bell state distributed to two distant parties without destroying it. The scheme requires two entangled states that are pre-shared between the parties, and we show that without these ancillary resources, the probability of non-destructively discriminating the Bell state is bounded by 1/4, which is the same as random guessing. Furthermore, we demonstrate a proof-of-principle experiment through an IonQ quantum computer that our scheme can surpass classical bounds when applied to practical quantum processor.
\end{abstract}

\maketitle

\section{Introduction} \label{chapter-introduction}

Discriminating different quantum states is at the heart of many quantum information applications. For example, the security of quantum key distribution is based on the difficulty of discriminating non-orthogonal quantum states~ \cite{bennett_quantum_2014, pirandola_advances_B020, xu_secure_2020}. It is noteworthy that B92 protocol is essentially equivalent to the unambiguous quantum state discrimination between two non-orthogonal qubits~\cite{bennett_quantum_1992}. 

\begin{figure}[b]
\includegraphics[width=0.44\textwidth]{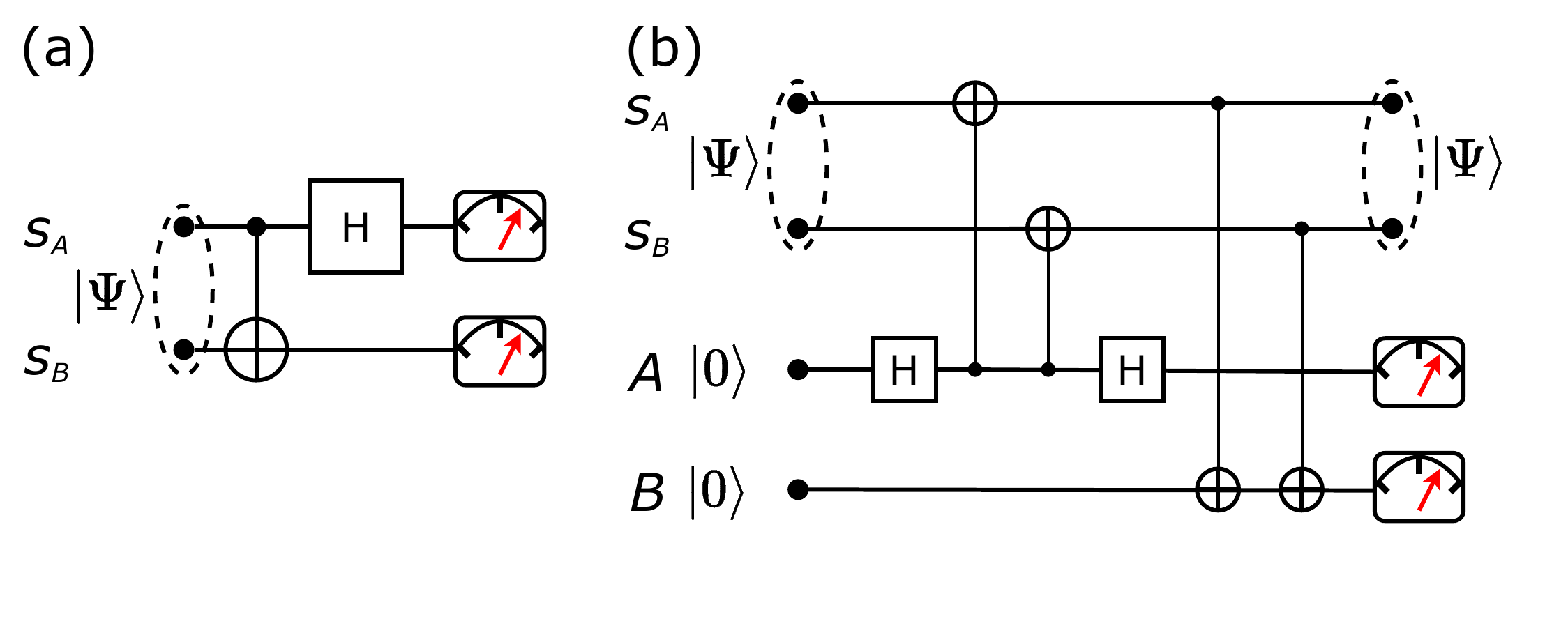}
\caption{Schemes for discriminating Bell states. (a) Direct measurement on the Bell state qubits. The quantum state is destroyed after knowing the state, thus it is unsuitable for further quantum information processing. (b) Nondestructive Bell state discrimination using ancillary qubits. The measurement result is registered by the outcome of the ancillary qubits and the Bell state qubits remain unchanged.}
\label{fig1}
\end{figure}

If we consider the case where a two-qubit system is prepared in one of the four Bell states, we can distinguish which Bell state it is in by using a circuit shown in Fig.~\ref{fig1}(a). This is known as Bell state measurement and is quintessential for measurement-device-independent quantum communications~\cite{braunstein_side-channel-free_2012, lo_measurement-device-independent_2012, choi_plug-and-play_2016, park_practical_2018}, quantum teleportation~\cite{photonics_teleportation}, and entanglement swapping~\cite{photonics_swapping}. However, this is a destructive measurement of the state. Therefore, if we have an arbitrary state and need to perform further quantum information processing after confirming what state it is, or if we desire to verify and validate it during process, we need a way to maintain the state without changing it even after distinguishing it. This can be achieved by utilizing ancillary qubits that interact with system qubits. For instance, as shown in Fig.~\ref{fig1}(b), Bell states of system qubits $s_A$ and $s_B$ can be nondestructively discriminated by the measurement results of two ancillary qubits $a_1$ and $a_2$ ~\cite{panigrahi_gupta_circuits_2007, gupta_general_2007, sisodia_experimental_2017, welte_nondestructive_2021}.
 
Bell states are often distributed between two parties who cannot perform direct non-local quantum operations between them. Therefore, it is necessary to verify and guarantee the state in such processes. However, Fig.~\ref{fig1}(b) scheme is not applicable to distant parties because it is necessary to implement $CNOT$ operations between an ancillary qubit with both Bell state qubits that are apart. Specifically, many quantum network applications begin with shared entangled particles $s_A$ and $s_B$ among distant parties. Since entanglement cannot be generated via local operation and classical communication (LOCC), entangled particles are usually generated by a {\it trusted} party and distributed via quantum channels~\cite{horodecki2009}. While the trusted third party is widely presumed in quantum network scenarios, a malicious third party or insecure quantum channels can cause significant threats in quantum communications~\cite{kim_informationally_2018, pramanik_equitable_2020}. Therefore, once entangled particles are shared by distant parties, it is desirable to check them in a nondestructive manner. 

Recently, it has been investigated as {\it quantum delocalized interaction} and shown that entangled ancillary qubits between parties are essential for nondestructive quantum state discrimination~\cite{paige_quantum_2020, lim_2023}. In this paper, we further investigate nondestructive quantum state discrimination of all four Bell states between distant parties. We theoretically investigate the upper bound of success probability with LOCC. Then, we present a scheme to discriminate all four Bell states without destroying it with pre-shared ancillary entangled qubits. We also demonstrate a proof-of-principle experiment using an IonQ quantum computer~\cite{ionq_site}. The experimental results verify the effectiveness of our protocol by showing that the experimental success probability surpasses the classical upper bound. 

\section{Theory} \label{chapter-theory}

\subsection{Nondestructive Bell state discrimination with local operations and classical communications}

Consider two distant parties, Alice and Bob, each receiving one qubit from a pair of qubits $|\Psi\rangle_{AB}$. Here, $|\Psi\rangle_{AB}$ is prepared in one of four Bell states, $|\phi^{\pm}\rangle=\frac{1}{\sqrt{2}}\left(|00\rangle\pm|11\rangle\right)$ and $|\psi^{\pm}\rangle=\frac{1}{\sqrt{2}}\left(|01\rangle\pm|10\rangle\right)$. Alice and Bob want to determine the Bell state $|\Psi\rangle_{AB}$ without altering the state. Now we refer the probability to successfully discriminate a given Bell state as $P_D$, the probability to a given Bell state remain unchanged after probe it as $P_F$.  The overall success probability for both cases would be $P_F P_D\le P_{\rm succ}\le {\rm min}\{P_F,P_D\}$. The lower bound is reached when $P_F$ and $P_D$ are independent, and the upper bound happens when $P_F$ and $P_D$ are highly correlated. 


A trivial strategy that Alice and Bob can take into account is random guessing. Alice and Bob do not touch the target Bell state at all while they just guess without any information. In this case, Bell state is unchanged so the nondestructive probability $P_F=1$. However, the probability to successful discrimination becomes $P_D=1/4$ since they randomly guess the state out of four possible Bell states. Therefore, the overall success probability of nondestructive quantum state discrimination becomes $P_{\rm succ}=1/4$. 
Another strategy can be performing simple projective measurements and reconstructing a Bell state from the measurement results. Although Alice and Bob cannot perform joint measurements, they can still perform projective measurements on their own qubits and obtain some information. Consider the case both Alice and Bob measure the Bell state in $Z$ basis and get the measurement result of $|0\rangle$ and $|0\rangle$ each. Then, they can assure that the state was in either $|\phi^+\rangle$ or $|\phi^-\rangle$. Therefore, the success probability for state discrimination becomes $P_D=1/2$. But even after Alice and Bob know their own qubit outcomes, they cannot recover an entangled state since they are separated. If they simply prepare the same quantum state from the outcome they achieved i.e., $|00\rangle$, nondestructive probability, or state overlap, with the original quantum state is $P_F=|\langle\phi^{\pm}|00\rangle|^2=1/2$. Therefore, the overall success probability of nondestructive quantum state discrimination becomes $P_{\rm succ}=1/4$. Actually, the success probability $P_{\rm succ}=1/4$ is the upper bound that Alice and Bob can achieve classically without genuine quantum resource i.e. ancillary entanglements.

\begin{theorem}
\label{theorem1}
The success probability to nondestructively discriminate all four Bell states distributed between two distant parties with local operations and classical communications (LOCC) can be at most $P_{\rm cl}=1/4$.
\end{theorem}

The proof of Theorem~\ref{theorem1} is represented in Appendix~\ref{a:Theorem1} in detail.

\subsection{Nondestructive Bell state discrimination with ancillary entanglement}

Having established that it is impossible to nondestructively discriminate four Bell states between distant parties without quantum resources. Now, we propose a pre-shared entanglement assisted scheme for nondestructive discrimination of Bell states. Note that the pre-shared entanglement can be considered as quantum resources and no further global operations between distant parties are necessary. Therefore, once the distant parties share ancillary entanglement, they can nondestructively discriminate Bell states via LOCC later on.

\begin{figure}[t]
\center{
\includegraphics[width=0.45\textwidth]{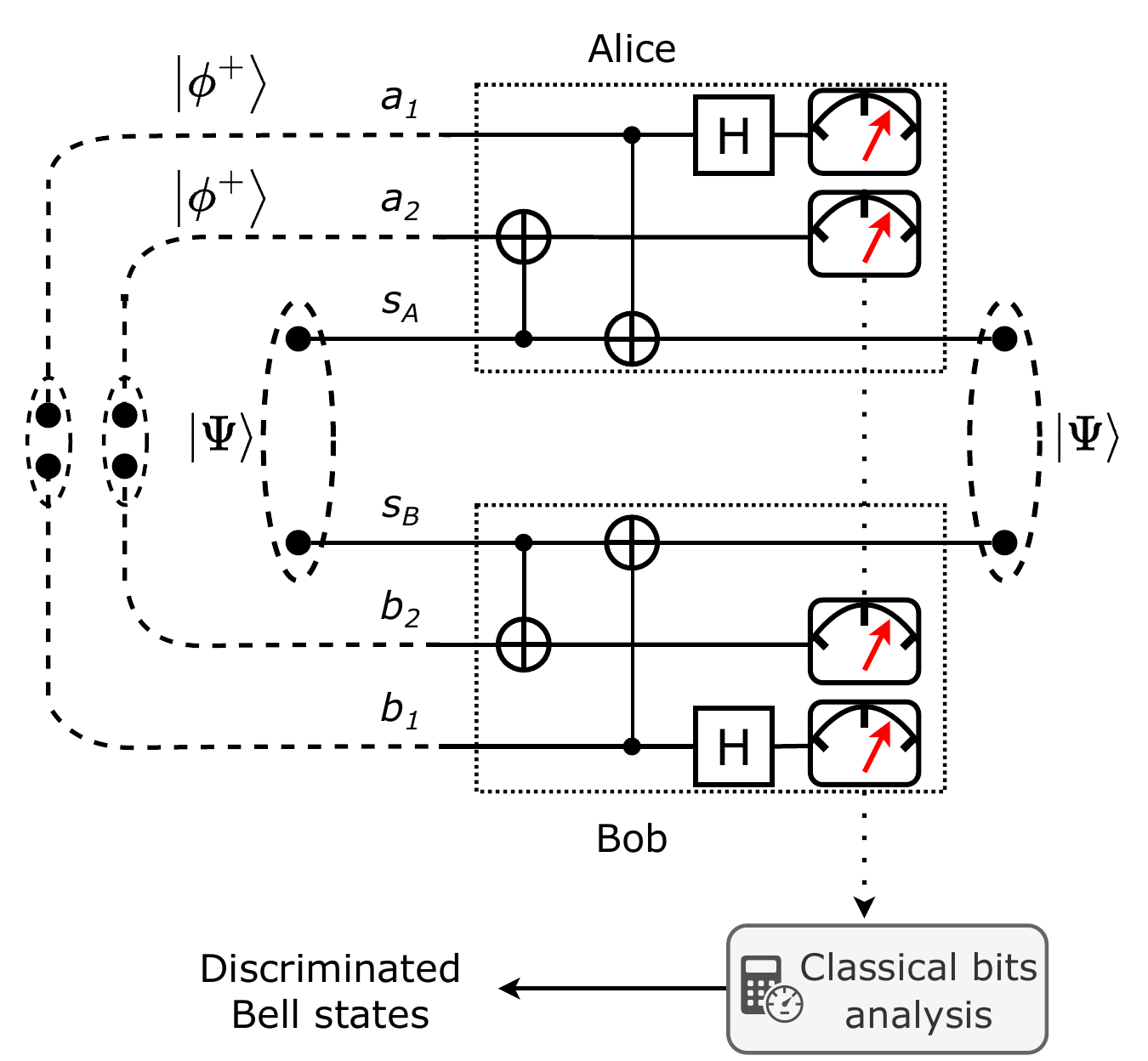}
\caption{Scheme for nondestructive Bell state discrimination between distant parties. The qubits $s_A$ and $s_B$ are prepared in one of four Bell states. This can be nondestructively discriminated with the help of two ancillary entanglement pairs $|\phi^+\rangle_{a_1b_1}$ and $|\phi^+\rangle_{a_2b_2}$.}
\label{scheme}
}
\end{figure}

Figure~\ref{scheme} shows our proposed scheme. It begins with two ancillary entangled states between Alice and Bob along with Bell state $|\Psi\rangle_{s_As_B}$ as follow.
\begin{eqnarray}
|\Psi\rangle_{\rm tot}&=&|\phi^+\rangle_{a_1b_1}\otimes|\phi^+\rangle_{a_2b_2}\otimes|\Psi\rangle_{s_As_B},
\label{initial}
\end{eqnarray}
where two ancillary entangled qubits are prepared in $|\phi^+\rangle$ and the target Bell state is in one of four Bell states, $|\Psi\rangle_{s_As_B}\in\{|\phi^{\pm}\rangle,|\psi^{\pm}\rangle\}$.

Let us assume that the target Bell state is prepared in $|\psi^-\rangle_{s_As_B}$. Then the state evolves as follows:
\begin{eqnarray}
|\Psi\rangle_{\rm tot,|\psi^-\rangle}=&&|\phi^+\rangle_{a_1b_1}\otimes|\phi^+\rangle_{a_2b_2}\otimes|\psi^-\rangle_{s_As_B}\nonumber\\
\xrightarrow{\phantom{a}{\rm X}_{s_Aa_2}\otimes{\rm X}_{s_Bb_2}\phantom{a}}&&|\phi^+\rangle_{a_1b_1}\otimes|\psi^+\rangle_{a_2b_2}\otimes|\psi^-\rangle_{s_As_B}\nonumber\\
\xrightarrow{\phantom{a}{\rm X}_{a_1s_A}\otimes{\rm X}_{b_1s_B}\phantom{a}}&&|\phi^-\rangle_{a_1b_1}\otimes|\psi^+\rangle_{a_2b_2}\otimes|\psi^-\rangle_{s_As_B}\nonumber\\
\xrightarrow{\phantom{aaa}{\rm H}_{a_1}\otimes{\rm H}_{b_1}\phantom{aaa}}&&|\psi^+\rangle_{a_1b_1}\otimes|\psi^+\rangle_{a_2b_2}\otimes|\psi^-\rangle_{s_As_B},
\nonumber
\label{initial}
\end{eqnarray}
where $X_{kl}$ and $H_{m}$ are a CNOT gate on $k$ (control) and $l$ (target) qubits and a Hadamard gate on qubit $m$, respectively. Note that all the CNOT gates are locally performed within Alice and Bob. Since both resultant ancillary pairs $a_1b_1$ and $a_2b_2$ are in $|\psi^+\rangle$, measuring them yields one of the outcomes $\{0101,\,0110,\,1001,\,1010\}$. Note that the Bell state $|\psi^-\rangle_{s_As_B}$ has not been changed during the procedure.
Similarly, one can find the state evolution with other Bell states as follow:
 
\begin{eqnarray}
|\Psi\rangle_{\rm tot,|\psi^+\rangle}\longrightarrow&&|\phi^+\rangle_{a_1b_1}\otimes|\psi^+\rangle_{a_2b_2}\otimes|\psi^+\rangle_{s_As_B},
\nonumber\\
|\Psi\rangle_{\rm tot,|\phi^-\rangle}\longrightarrow&&|\psi^+\rangle_{a_1b_1}\otimes|\phi^+\rangle_{a_2b_2}\otimes|\phi^-\rangle_{s_As_B},
\nonumber\\
|\Psi\rangle_{\rm tot,|\phi^+\rangle}\longrightarrow&&|\phi^+\rangle_{a_1b_1}\otimes|\phi^+\rangle_{a_2b_2}\otimes|\phi^+\rangle_{s_As_B}.\nonumber
\end{eqnarray}

It is straightforward to see that the different Bell states $|\Psi\rangle_{s_As_B}$ provide distinct ancillary qubit states while the Bell states are unaltered. Therefore, the Bell state can be {\it completely} nondestructively discriminated by measuring the ancillary qubits. Table~\ref{table-bits} summarizes the measurement results of the ancillary qubits according to the Bell states. 

\begin{table}[b]
\begin{tabular}{c|c} 
  \hline
  ~~~~~$|\Psi\rangle_{s_As_B}$~~~~~ & ~~~~~~~~~Bit values ($a_1b_1a_2b_2$)~~~~~~~~~ \\
  \hline
  $\ket{\phi^+}_{s_As_B}$ & 0000, 0011, 1100, 1111  \\ 
  \hline
  $\ket{\phi^-}_{s_As_B}$ & 0111, 0100, 1000, 1011  \\ 
    \hline
  $\ket{\psi^+}_{s_As_B}$ & 0001, 0010, 1110, 1101 \\ 
  \hline
  $\ket{\psi^-}_{s_As_B}$ & 0101, 0110, 1001, 1010  \\ 
  \hline
\end{tabular}
\caption{Measurement results of ancillary qubits according to the Bell state $|\Psi\rangle_{s_As_B}$. Both ancillary entanglement pairs are prepared in $|\phi^+\rangle_{ab}$.}
\label{table-bits}
\end{table}

\subsection{Nondestructive Bell state discrimination with non-maximally entangled ancilla}
\label{subchapter_c}

In the previous section, we considered the case where both ancillary qubits were in the $|\phi^+\rangle$, which is the maximally entangled. In this section, we explore the case where ancillary qubits are not in maximally entangled states. Consider the case of analyzing the target Bell state by utilizing the Werner state, an incoherent combination of a pure maximally entangled state and a completely mixed state. The Werner state is written as
\begin{eqnarray}
W_2&=&(1-\lambda)|\phi^+\rangle \langle \phi^+| + \lambda \frac{\mathbb{I}}{4},
\label{werner_single}
\end{eqnarray}
where $\lambda$ determines depolarizing noise added in maximally entangled state $|\phi^+\rangle$. When $\lambda=0$, the $W_2$ becomes a density matrix of maximally entangled state $|\phi^+\rangle$, and $\lambda=1$ makes a state maximally mixed state. 
Note that $\mathbb{I}$ can be rewritten in terms of mixture of all four-Bell states, and thus, the Werner state can be rewritten as
\begin{eqnarray}
W_2&=& (1-\frac{3 \lambda}{4}) |\phi^+\rangle \langle \phi^+| + \frac{\lambda}{4}(|\phi^-\rangle \langle \phi^-| \nonumber \\ && +|\psi^+\rangle \langle \psi^+|+|\psi^-\rangle \langle \psi^-|).
\label{werner_single2}
\end{eqnarray}


The only term we need to focus on is $|\phi^+\rangle\langle\phi^+|$, which is the Bell basis term that can be applied to nondestructive Bell state discrimination we are considering. It is worth noting that the arbitrary maximally entangled states can be used to achieve non-destructive discrimination, but with different output combination we have to know in initial. Here we are only considering $|\phi^+\rangle\langle \phi^+|$ as an example. 

In case of two pairs of ancilla, Eq.~(\ref{werner_single}), the state can be represented by a product state of two Werner states $W_2 \otimes W_2$. 
Then, total ancillary state can be written as follow:
\begin{eqnarray}
\label{werner_double}
W_2 \otimes W_2 &=&(1-\frac{3 \lambda}{4})^2|\phi^+ \phi^+\rangle \langle \phi^+ \phi^+|  \\
&+& (1-\frac{3 \lambda}{4})(\frac{\lambda}{4}) \{{\rm terms ~ with ~ one ~ |\phi^+\rangle} \}  \nonumber \\
&+& (\frac{\lambda}{4})^2 \{{\rm terms ~ with ~ no ~ |\phi^+\rangle} \}, \nonumber   
\end{eqnarray}
and an entire equation would be revealed in Appendix~\ref{a:twowerner}. Here we assume that both ancilla has same value of $\lambda$.

\begin{figure}[t]
\includegraphics[width=0.43\textwidth]{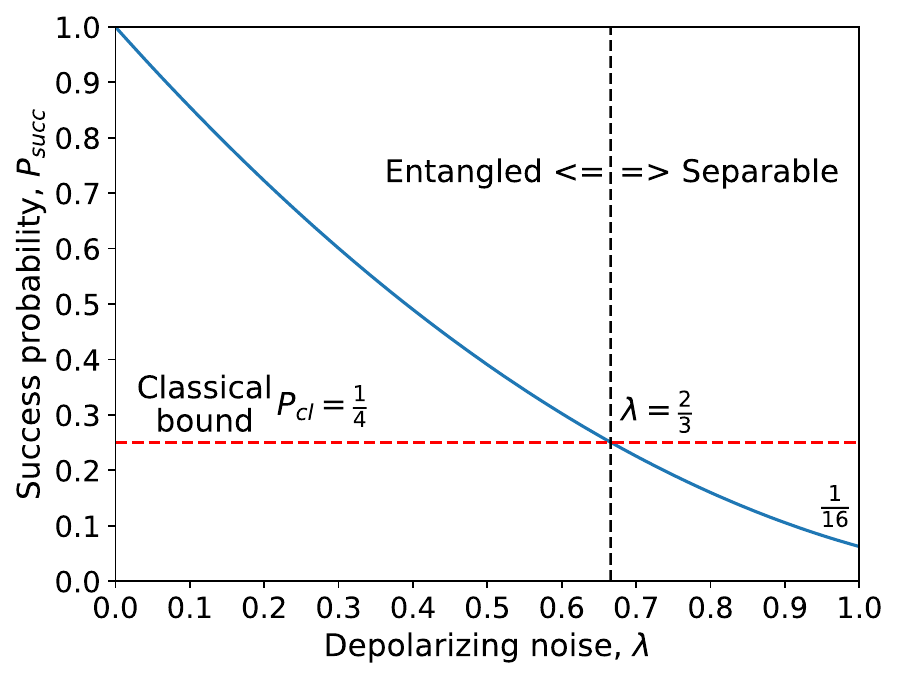}
\caption{Success probability depends on value of $\lambda$. Analytical solution is based on expanded for two ancilla Werner state. When $\lambda=0$, the Werner state becomes two pairs of $|\phi^+\rangle$, then success probability would be 1. However, as increasing $\lambda$, the result goes to $\frac{1}{16}$ since the ancillary Werner state becomes maximally mixed one.}
\label{analytical}
\end{figure}

It is straightforward to see that only the case $|\phi^+ \phi^+ \rangle \langle \phi^+ \phi^+|$ can successfully discriminate a Bell state without destroying it, and it appears with probability $P_{\rm succ}=(1-\frac{3}{4} \lambda)^2$. Otherwise, the estimation of the state would be incorrect, and/or the determination would contaminate the target Bell state. Now, we can plot the success probability $P_{\rm succ}$ of our protocol with respect to $\lambda$ as shown in Fig.~\ref{analytical}. For the case $\lambda = \frac{2}{3}$, the Werner state with no entanglement, cause the success probability to lose the quantum advantage over the classical one, $P_{\rm cl}=\frac{1}{4}$. And when the ancilla is maximally mixed state with $\lambda=1$, the success probability becomes $P_{\rm succ}=\frac{1}{16}$, which is the case the target Bell state is contaminated from the original one, and the discrimination becomes random.

\section{Experiment using IonQ} \label{chapter-ionq}

\subsection{Results for nondestructive Bell state discrimination with maximally entangled ancilla}

\begin{figure}[t]
\center{
\includegraphics[width=0.44\textwidth]{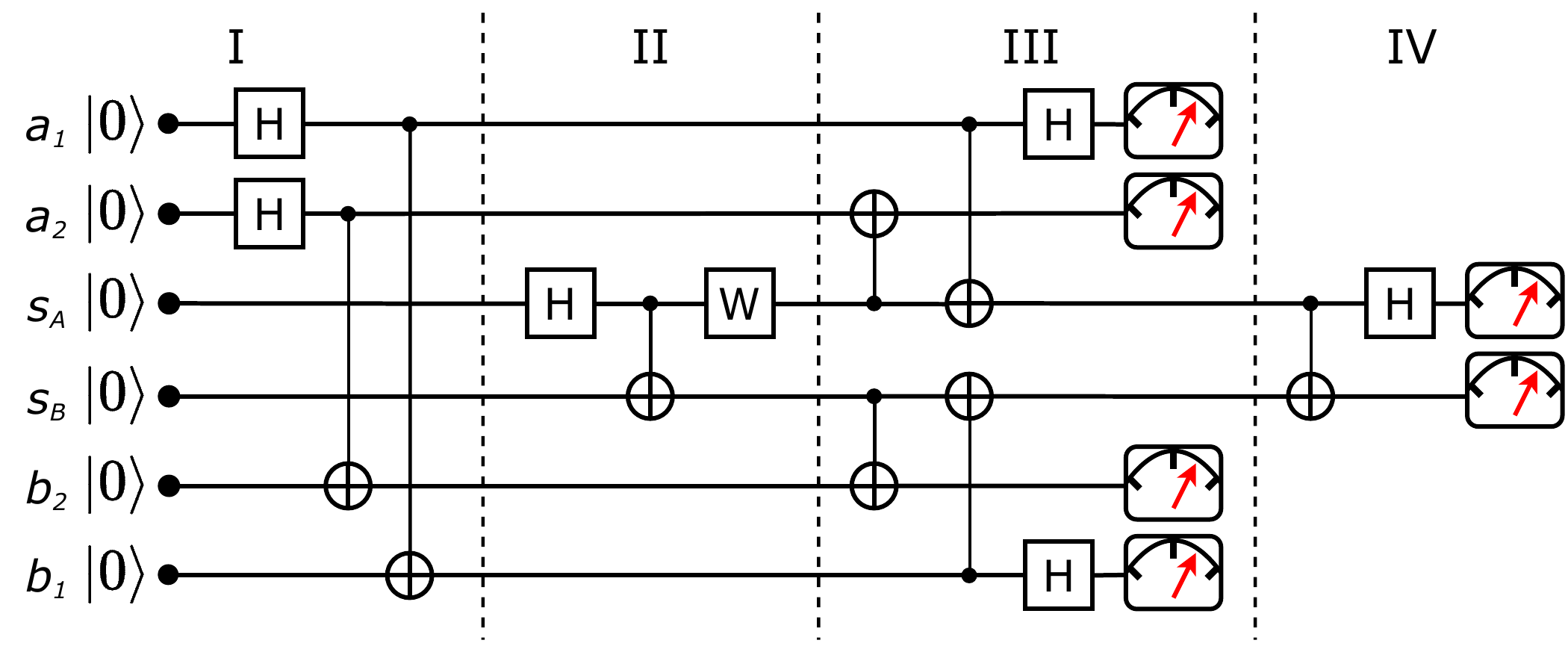}
\caption{A complete quantum circuit to perform a proof-of-principle experiment on IonQ quantum computer. It consists of 4-steps described in main text. Here, $W$ is a single qubit Pauli gate where $W\in\{I,X,Y,Z\}$ to prepare arbitrary Bell state.}
\label{IonQ}
}
\end{figure}

We demonstrate our protocol using an IonQ quantum computer~\cite{ionq_site}. Figure~\ref{IonQ} shows the quantum circuit run on the IonQ quantum computer. The entire quantum circuit is composed of four sections as follow:

\phantom{aa}

\noindent{\rm I. Ancillary entanglement preparation--.} 
By applying Hadamard gate and CNOT gate to two pairs of two ancillary qubits, we generate two pairs of entangled states $|\phi^+\rangle_{a_1b_1}$ and $|\phi^+\rangle_{a_2b_2}$. 

\phantom{aa}

\noindent{\rm II. Preparation of target Bell state--.} The Bell state to probe is prepared in one of four Bell states. Arbitrary Bell states $|\Psi\rangle$ are prepared with different single qubit gates $W\in\{I,X,Y,Z\}$, where $I$ is an identity operator and $X, Y,$ and $Z$ are the Pauli operators.

\phantom{aa}

\noindent{\rm III. Interaction with ancillary qubits--.} The interactions between ancillary states and target Bell state are performed to implement nondestructive entanglement discrimination. The quantum circuit is identical to that of Fig.~\ref{scheme}. Note that no direct non-local operations between Alice and Bob are allowed as they are assumed to be separate. The Bell state discrimination result is registered as bit values of $a_1b_1a_2b_2$, refer Table~\ref{table-bits}.

\phantom{aa}

\noindent{\rm IV. Verifying non-destructive characteristic--.} We perform a Bell state measurement to check whether the Bell state prepared in step II has been altered or not. This is identical to that of Fig.~\ref{fig1}(a). Note that this is unnecessary for other applications.

\phantom{aa}

\begin{figure}[t]
\center{
\includegraphics[width=0.43\textwidth]{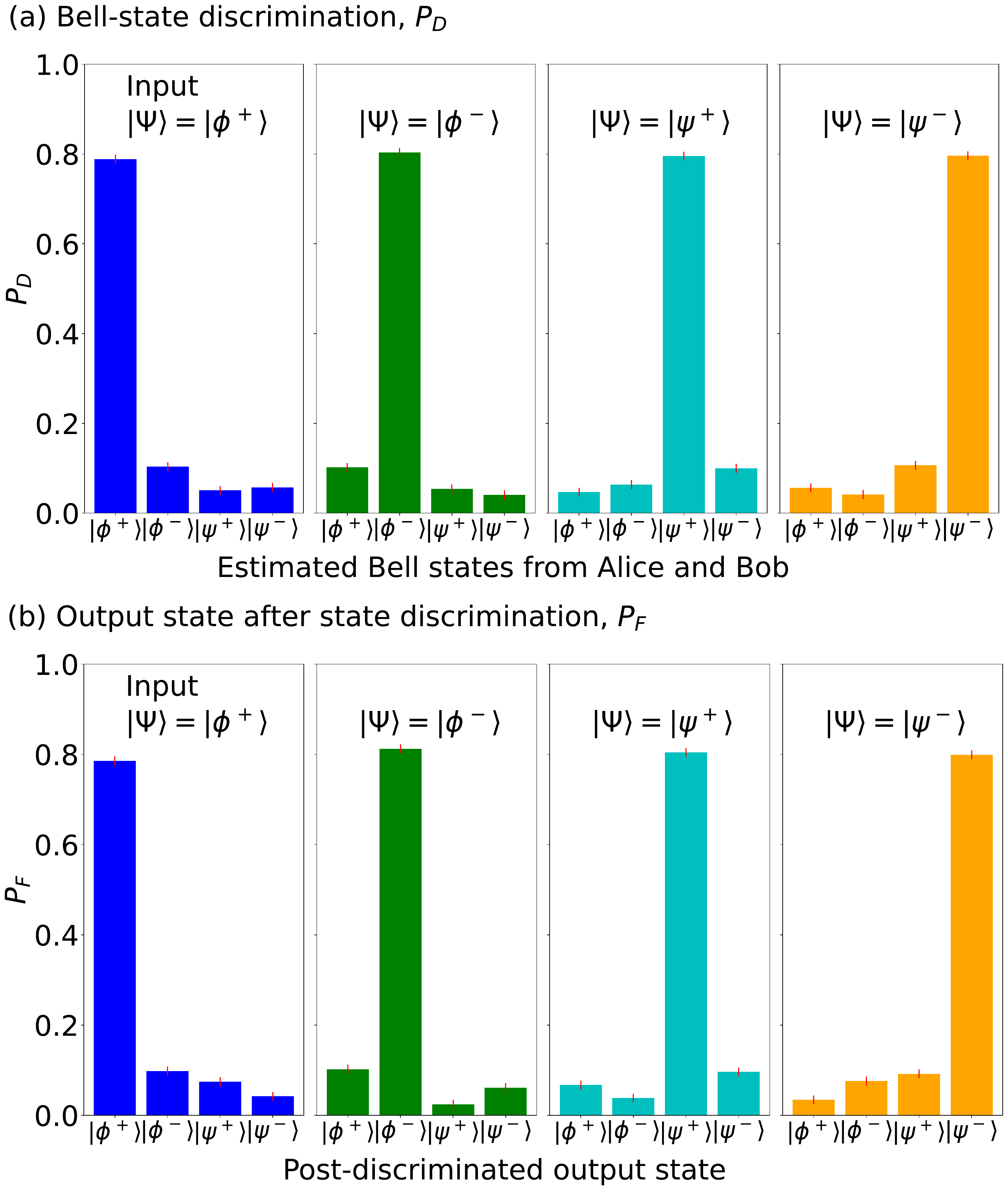}
\caption{Experimental results using the IonQ quantum computer. (a) Bell state discrimination probability given by the measurement results of ancillary qubits. (b) The probability that Bell state are measured in Bell bases after the discrimination.}
\label{bsd-and-nd}
}
\end{figure}

In our experiment, we repeat 10,000 times of the experimental runs for each Bell state. The state discrimination results depending on the different Bell states are shown in Figure~\ref{bsd-and-nd}(a). It clearly shows that the Bell states are well discriminated with high probability, the average successful state discrimination probability $P_{D}=0.796\pm0.005$ for all four Bell states. Figure~\ref{bsd-and-nd}(b) shows the results of post-discriminate states. One can see the initial state has been maintained with high probability, the average probability that the Bell states remain unchanged is given as $P_{F}=0.800\pm0.010$.

\begin{figure}[t]
\includegraphics[width=0.44\textwidth]{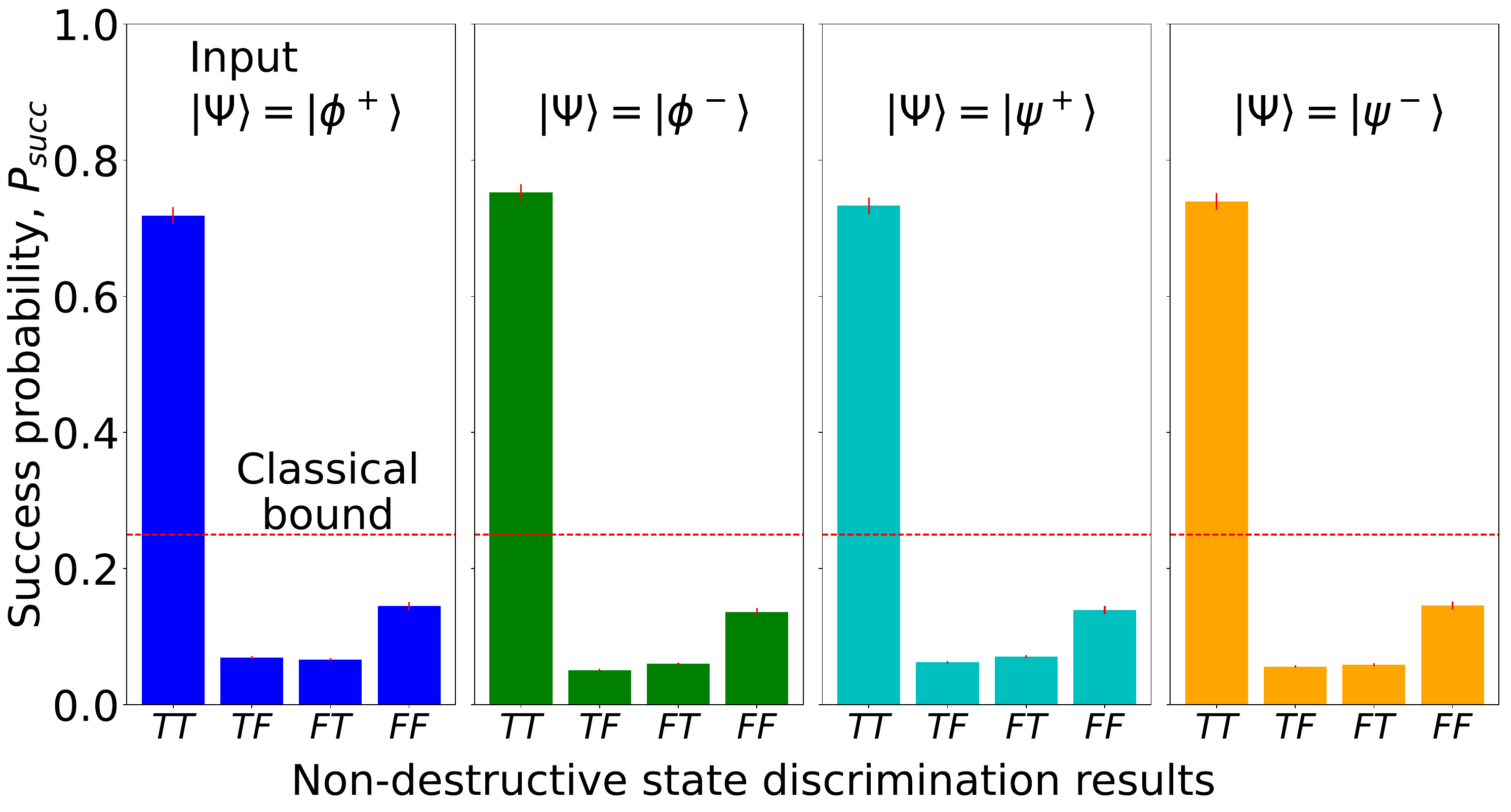}
\caption{Experimental truth table for nondestructive entanglement discrimination: i) successful state discrimination with unchanged Bell state (TT), ii) successful state discrimination but Bell state altered (TF), iii) Failed to discriminate the state but Bell state are unchanged (FT), and iv) state discrimination failed and Bell state altered (FF).}
\label{truth-table}
\end{figure}

The complete experimental results can be categorized as a following truth table: i) successful state discrimination with unchanged Bell state (TT), ii) successful state discrimination but Bell state altered (TF), iii) Failed to discriminate the state but Bell state is unchanged (FT), and iv) state discrimination failed and Bell state is altered (FF). The successful nondestructive Bell states discrimination result corresponds to simultaneous success of state discrimination and unaltered Bell state, i.e., TT case. Figure~\ref{truth-table} presents the truth table with the prepared Bell states. The nondestructive discrimination probability averaged over for all four Bell states is $P_{\rm succ}=0.736\pm0.012$. The success probability $P_{\rm succ}$ achieved by the practical quantum computer is well above the upper bound of Bell state discrimination without shared entanglements, $P_{\rm cl}=1/4$. Note that $P_{\rm succ}\le P_DP_F$ means there is a correlation between $P_D$ and $P_F$.  We also can find such correlations, as shown in Fig.~\ref{truth-table}, the probability for FF cases is much higher than individual failures of state discrimination (FT) and non-destructiveness (TF).


\subsection{Results for nondestructive Bell state discrimination with non-maximal entanglements}

\begin{figure}[h]
\includegraphics[width=0.43\textwidth]{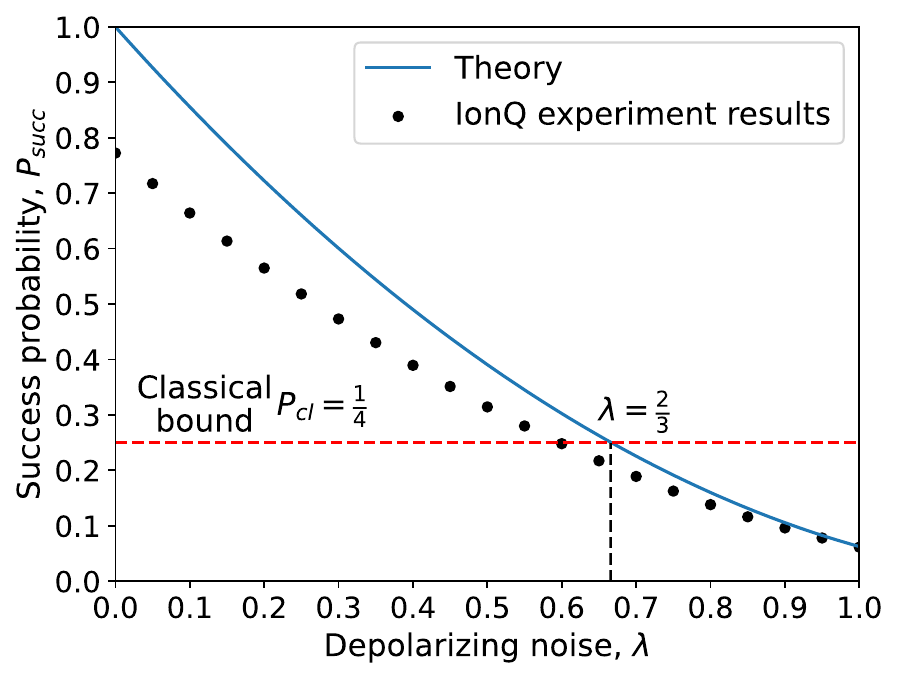}
\caption{Success probability $P_{\rm succ}$ depends on value of $\lambda$ for two ancilla Werner states. Theoretical result (blue curve) is drawn for comparison. The blue curve shows the theoretical prediction of $P_{\rm succ}$ based on the $\lambda$, analyzed in Sec.~II-C. Each point of IonQ experimental result is produced by taking 9600 shots, and repeated 100 times. The deviations are small enough to reside inside the dots. Even when $\lambda=0$, the IonQ result does not reach $P_{\rm succ}=1$ due to the noises of the IonQ processor. As $\lambda$ increases, the result reaches $\frac{1}{16}$, which is predicted in the theory, according to the maximally mixed ancilla.}
\label{analytical_ionq}
\end{figure}

In this section we show IonQ experiment results with Werner states ancilla, which are analyzed in Chapter~\ref{subchapter_c}. We dealt with the statistics of outcomes of events that can occur in Werner states ancilla, similar to twirling technique, to explore the affect by Werner states ancilla with arbitrary $\lambda$ values. Specifically, since it is impractical to realize Werner states with arbitrary $\lambda$ as an input for practical quantum computer, we run the quantum computer with all the states composing Werner state, and see the statistical combinations of these outcomes according to the weights derived from $\lambda$.
The results are implemented only for the case discriminating Bell $|\phi^+ \rangle$ state. However, the results can be extended to other Bell states without loss of generality.
The results of analytic analysis in Chapter~\ref{subchapter_c} and its implementation by IonQ computer is shown on Fig.~\ref{analytical_ionq}. For each Werner state, Eq.~\eqref{werner_double}, with different $\lambda$, which are represented as black circles in Fig.~\ref{analytical_ionq}, we run circuit with 9600 shots, and repeat the process 100 times. The standard deviations for each point are less than 0.5\% and smaller than the size of markers. We can see the entangled ancillary state with $\lambda$ that is larger than 0.6, overcome classical bound which is close to theoretical bound $\lambda=\frac{2}{3}$. 


\section{Conclusions}

While many quantum communication and network protocols begin with pre-shared entanglement between distant parties, it could have been contaminated during the transmission or by the malicious third party. Therefore, it is of importance to verify the entanglement before the quantum communications. Since the entanglement will be utilized for further quantum information processing, the verification should be done in a nondestructive manner. 

In this paper, we have shown that the success probability of nondestructive quantum state discrimination of all four Bell states can be at most $P_{\rm cl}=1/4$ with local operations and classical communications. The limited success probability can be surpassed by utilizing pre-shared entanglement between parties. In particular, we have proposed a scheme for complete nondestructive Bell state discrimination using two ancillary entangled qubit pairs and verify the experimental feasibility by demonstrating a proof-of-principle experiment on an IonQ quantum computer. It would be interesting to extend our scheme to discriminate entanglement of high dimensional and/or multipartite quantum states.

\section{Acknowledgements}
\noindent This research was funded by Korea Institute of Science and Technology (2E32241, 2E32801), National Research Foundation of Korea (2023M3K5A1094805, 2023M3K5A1094805) and Institute for Information \& communications Technology Planning \& Evaluation (IITP) (RS-2023-00222863, 2022-0-00463). H.K. is supported by the KIAS Individual Grant No. CG085301 at Korea Institute for Advanced Study.

\appendix
\section{Proof of Theorem 1} 
\label{a:Theorem1}
Here, we prove \textbf{Theorem 1.} proposed in main manuscript, which is bound the success probability of nondestructive discrimination of all four Bell states distributed between two distant parties under LOCC condition.
\begin{proof}
Consider the case a quantum game in which Alice and Bob receive an arbitrary Bell state and guess which Bell basis it is in without altering it. We define a probability $p_{win}$ to win  this game as a probability to discriminate the Bell state and simultaneously to remain Bell state unchanged, and average over for all possible Bell states. Then, we can represent the situation as an equation below
\begin{widetext}
\begin{equation}
p_{win} = \sum_\Psi \sum_{\lambda_\Psi} p_\Psi \mbox{Tr} \Big{[} \Pi^{\lambda_\Psi} _{A'B'} \otimes |\Psi\rangle \langle \Psi|_{AB} \big{(} W (\rho_{A'B'} \otimes |\Psi\rangle \langle \Psi|_{AB}) W^\dagger \big{)} \Big{]}.
\label{winprob}
\end{equation}
\end{widetext}
Here, $|\Psi\rangle_{AB} = \{ |\psi^\pm\rangle,|\phi^\pm \rangle \} $, the Bell state Alice and Bob want to discriminate and the subscript $AB$ means that the Bell state is distributed to Alice and Bob, respectively. So the indices $A$ and $B$ indicate Bell state of Alice's and Bob's side, respectively, $A'$ and $B'$ are for their ancillary system $\rho_{A'B'} = |z\rangle \langle z|_{A'B'}$ to interact with the Bell state $|\Psi\rangle_{AB}$. $W = U_{AA'} \otimes V_{BB'}$ means a composition of local unitary operations $U_{AA'}$ and $V_{BB'}$, and each means the local unitary between $A$ and $A'$ and the local unitary between $B$ and $B'$, respectively. And $\Pi^{\lambda_\Psi} _{A'B'} = |\lambda_\Psi\rangle \langle \lambda_\Psi|_{A'B'}$ is the measurement operators with outcome $\lambda_\Psi$, which give indicators for each possible Bell state. As an example, $\lambda_z$ can be $00,11,01,10$ and each two-bit string correspond to Bell states of four cases revealed from measuring $\Pi_{A'B'}$. For simplicity, rewrite the Eq.~(\ref{winprob}) in a shorter form with Kraus operator~\cite{qitd412} $K^{\lambda_\Psi} _{AB} =\ _{A'B'}\langle \lambda_\Psi |W|z\rangle_{A'B'}$, which defines a Kraus operator that starts from ancillary system $|z\rangle_{A'B'}$ and apply local operations $W$, then projects to $|\lambda_\Psi \rangle_{A'B'}$. Then, Eq.~\eqref{winprob} becomes
\begin{equation}
	p_{win} = \frac{1}{4} \sum_\Psi \sum_{\lambda_\Psi} \Big{|}\  _{AB}\langle \Psi| K^{\lambda_\Psi} _{AB}|\Psi\rangle_{AB} \Big{|}^2.
\label{winprobK}
\end{equation}
Here, we assume that the Bell states are given with equal probability $p_{|\Psi\rangle}=1/4$ for all possible bases, and the winning probability determines the probability to given Bell state $|\Psi\rangle_{AB}$ remains itself after the Kraus operator $K^{\lambda_\Psi}_{AB}$, for every possible Bell state and all possible indicators $\lambda_{\Psi}$.

As we consider the case that Alice and Bob cannot share entanglement i.e., $\rho_{A'B'}$ is a separable state, and the measurements $\Pi^{\lambda_\Psi}_{A'B'}$ is restricted to be local. Then, Kraus operator for whole process can be decomposed into Alice's and Bob's part independently, and Eq.~\eqref{winprobK} can be represented as
\begin{equation}
		p_{win} = \frac{1}{4} \sum_\Psi \sum_{\lambda_\Psi} \Big{|}\  _{AB}\langle \Psi| K^{\lambda_\Psi} _{A} \otimes K^{\lambda_\Psi} _{B}|\Psi\rangle_{AB} \Big{|}^2.
\label{winprobL}
\end{equation}
Here, we remove the subscript $AB$ in $|\Psi\rangle$ for better readability. Then, the detailed calculation for the bound of $p_{win}$ under LOCC is as follows:

\begin{widetext}
\begin{eqnarray*}
p_{win} &=& \frac{1}{4} \sum_{\Psi} \sum_{\lambda_\Psi}  {\Big{|}\langle \Psi|K^{\lambda_\Psi}_A \otimes K^{\lambda_\Psi}_B|\Psi\rangle \Big{|}}^2 \\
&=& \frac{1}{4} \sum_{\Psi}\sum_{\lambda_\Psi} \mbox{Tr} {\Big{[}|\Psi\rangle \langle \Psi |{K^{\lambda_\Psi}_A}^\dagger \otimes {K^{\lambda_\Psi}_B}^\dagger|\Psi\rangle \langle \Psi|K^{\lambda_\Psi}_A \otimes K^{\lambda_\Psi}_B\Big{]}} \\
&=& \frac{1}{4} \sum_{\Psi}\sum_{\lambda_\Psi} \mbox{Tr} {\Big{[}|\Psi\rangle \langle \Psi |({K^{\lambda_\Psi}_A}^\dagger \otimes \mathbb{I}_B) (\mathbb{I}_A \otimes  {K^{\lambda_\Psi}_B}^\dagger)|\Psi\rangle \langle \Psi|(K^{\lambda_\Psi}_A \otimes \mathbb{I}_B) (\mathbb{I}_A \otimes  K^{\lambda_\Psi}_B)}\Big{]} \\
&=& \frac{1}{4} \sum_{\Psi}\sum_{\lambda_\Psi} \mbox{Tr} {\Big{[} ({K^{\lambda_\Psi}_A}^\dagger \otimes \mathbb{I}_B) |\Psi\rangle \langle \Psi | (K^{\lambda_\Psi}_A \otimes \mathbb{I}_B)  (\mathbb{I}_A \otimes  K^{\lambda_\Psi}_B) |\Psi\rangle \langle \Psi | (\mathbb{I}_A \otimes  {K^{\lambda_\Psi}_B}^\dagger)}\Big{]} \\
&\le & \frac{1}{4} \sum_{\Psi}\sum_{\lambda_\Psi} \mbox{Tr} \Big{[}{ ({K^{\lambda_\Psi}_A}^\dagger \otimes \mathbb{I}_B) |\Psi\rangle \langle \Psi | (K^{\lambda_\Psi}_A \otimes \mathbb{I}_B)}\Big{]} \mbox{Tr}  \Big{[}{(\mathbb{I}_A \otimes  K^{\lambda_\Psi}_B) |\Psi\rangle \langle \Psi | (\mathbb{I}_A \otimes  {K^{\lambda_\Psi}_B}^\dagger)}\Big{]} \\
&=& \frac{1}{4} \sum_{\Psi}\sum_{\lambda_\Psi} \mbox{Tr}_A \Big{[}{K_A^{\lambda_\Psi} {K^{\lambda_\Psi}_A}^\dagger \mbox{Tr}_B (|\Psi \rangle \langle \Psi|)}\Big{]}  \mbox{Tr}_B \Big{[}{K^{\lambda_\Psi}_B}^\dagger K_B^{\lambda_\Psi} \mbox{Tr}_A (|\Psi \rangle \langle \Psi|)\Big{]} \\
&=& \frac{1}{4} \sum_{\Psi}\sum_{\lambda_\Psi} \mbox{Tr}_A \Big{[}{K_A^{\lambda_\Psi} {K^{\lambda_\Psi}_A}^\dagger \frac{\mathbb{I}_A}{2}}\Big{]}  \mbox{Tr}_B \Big{[}{K^{\lambda_\Psi}_B}^\dagger K_B^{\lambda_\Psi} \frac{\mathbb{I}_B}{2}\Big{]} \\
&=& \frac{1}{4} \sum_{\Psi}\sum_{\lambda_\Psi}  \frac{1}{4} \mbox{Tr} \Big{[}K_A^{\lambda_\Psi} {K^{\lambda_\Psi}_A}^\dagger \otimes {K^{\lambda_\Psi}_B}^\dagger K_B^{\lambda_\Psi} \Big{]} \\
&=& \frac{1}{16} \mbox{Tr}\big{[}\mathbb{I}_{AB}\big{]} = \frac{1}{4}  
\end{eqnarray*}
\end{widetext}
Here, $\mathbb{I}_i$ means identity operator for $i$, the relation that $\mbox{Tr}[AB] \le \mbox{Tr}[A]\mbox{Tr}[B]$ for $A, B \geq 0$ is used for the inequality~\cite{englert_inequality}. We also take into account the fact that a partial trace for one-side of any Bell state is maximally mixed state.\\
\end{proof}
\newpage
\section{Detailed representation of a pair of two-qubit Werner state}
\label{a:twowerner}
Here we show the state $W_2 \otimes W_2$, a pair of two-qubit Werner state, which is used as an ancillary state for our non-destructive Bell state discrimination protocol. Though the state is briefly described for brevity in the main text, it can be fully described as follows:

\begin{widetext}
\begin{eqnarray*}
W_2 \otimes W_2 &=&(1-\frac{3 \lambda}{4})^2|\phi^+ \phi^+\rangle \langle \phi^+ \phi^+|  \\
&+& (1-\frac{3 \lambda}{4})(\frac{\lambda}{4}) \Big{(} |\phi^+ \psi^+\rangle \langle \phi^+ \psi^+| + |\phi^+ \psi^-\rangle \langle \phi^+ \psi^-| + |\phi^+ \phi^-\rangle \langle \phi^+ \phi^-| \nonumber \\
&+&|\psi^+ \phi^+\rangle \langle \psi^+ \phi^+| + |\psi^- \phi^+\rangle \langle \psi^- \phi^+| + |\phi^- \phi^+\rangle \langle \phi^- \phi^+| \Big{)} \nonumber \\
&+& (\frac{\lambda}{4})^2 \Big{(} |\psi^+ \psi^+\rangle \langle \psi^+ \psi^+| + |\psi^+ \psi^-\rangle \langle \psi^+ \psi^-| + |\psi^+ \phi^-\rangle \langle \psi^+ \phi^-| + |\psi^- \psi^+\rangle \langle \psi^- \psi^+| \nonumber \\
&+& |\psi^- \psi^-\rangle \langle \psi^- \psi^-| + |\psi^- \phi^-\rangle \langle \psi^- \phi^-| + |\phi^- \psi^+\rangle \langle \phi^- \psi^+| + |\phi^- \psi^-\rangle \langle \phi^- \psi^-| + |\phi^- \phi^-\rangle \langle \phi^- \phi^-| \Big{)} \nonumber
\label{a:full}
\end{eqnarray*}
\end{widetext}

Only the term $|\phi^+ \phi^+\rangle$ makes non-destructive Bell state discrimination works, and other terms bring $P_F=0$, i.e. change target Bell state to other one, and/or $P_D=0$, i.e. fail to discriminate the target Bell state. All possible results summarized in the table~\ref{a:table-werner} below:

\begin{table}[h]
\begin{tabular}{c|c|c} 
  \hline
  basis of ancilla  & ~~~~$P_F$~~~~ & ~~~~$P_D$~~~~ \\
  \hline
  $\ket{\phi^+ \phi^+}$ & 1 & 1 \\ 
  \hline
  $\ket{\phi^+ \psi^+}$ & 1 & 0  \\ 
  \hline
  $\ket{\phi^+ \psi^-}$ & 0 & 0  \\ 
  \hline
  $\ket{\phi^+ \phi^-}$ & 0 & 0  \\  
  \hline
  $\ket{\psi^+ \phi^+}$ & 0 & 1 \\ 
  \hline
  $\ket{\psi^+ \psi^+}$ & 0 & 0  \\ 
  \hline
  $\ket{\psi^+ \psi^-}$ & 0 & 0  \\ 
  \hline
  $\ket{\psi^+ \phi^-}$ & 0 & 0  \\  
  \hline
  $\ket{\psi^- \phi^+}$ & 0 & 0 \\ 
  \hline
  $\ket{\psi^- \psi^+}$ & 0 & 0  \\ 
  \hline
  $\ket{\psi^- \psi^-}$ & 0 & 0  \\ 
  \hline
  $\ket{\psi^- \phi^-}$ & 0 & 1  \\  
  \hline
  $\ket{\phi^- \phi^+}$ & 1 & 0 \\ 
  \hline
  $\ket{\phi^- \psi^+}$ & 1 & 0  \\ 
  \hline
  $\ket{\phi^- \psi^-}$ & 0 & 0  \\ 
  \hline
  $\ket{\phi^- \phi^-}$ & 0 & 1  \\  
  \hline
\end{tabular}
\caption{All the events created by each basis of ancillary Werner state. Only the case $|\phi^+ \phi^+\rangle$ makes non-destructive discrimination of Bell state possible in our scenario.}
\label{a:table-werner}
\end{table}

\end{document}